\documentclass[preprint]{aastex}

\slugcomment{Accepted for publication, Astrophysical Journal,
Sept.\ 1, 2006\\\copyright 2006 The Astrophysical Journal\\
{\tt http://www.journals.uchicago.edu/ApJ/}}

\shorttitle{Electron Acceleration at Collisionless Shocks}
\shortauthors{Burgess}

\begin{document}


\title{Simulations of Electron Acceleration at Collisionless Shocks: The
Effects of Surface Fluctuations}


\author{D. Burgess}
\affil{Astronomy Unit, Queen Mary, University of London, London E1 4NS, UK}


\begin{abstract}
Energetic electrons are a common feature of interplanetary shocks and
planetary bow shocks, and they are invoked as a key component of models
of nonthermal radio emission, such as
solar radio bursts. A simulation study is carried out of electron acceleration
for high Mach number, quasi-perpendicular shocks, typical of the
shocks in the solar wind. Two dimensional self-consistent hybrid
shock simulations provide the electric and magnetic fields in which
test particle electrons are followed. A range of different shock types,
shock normal angles, and injection energies are studied.
When the Mach number is low, or the simulation configuration suppresses
fluctuations along the magnetic field direction, the results agree with
theory assuming magnetic moment conserving reflection (or Fast Fermi
acceleration), with electron energy gains of a factor only $2 - 3$.
For high Mach number, with a realistic simulation configuration, the
shock front has a dynamic rippled character. The corresponding
electron energization is radically different: Energy spectra display:
(1) considerably higher maximum energies than Fast Fermi acceleration;
(2) a plateau, or shallow sloped region, at intermediate energies
$2 - 5$ times the injection energy; (3) power law fall off with increasing
energy, for both upstream and downstream particles, with a slope
decreasing as the shock normal angle approaches perpendicular; (4) sustained
flux levels over a broader region of shock normal angle than for
adiabatic reflection. All these features are in good qualitative agreement
with observations, and show that dynamic structure in the shock surface
at ion scales produces effective scattering
and can be responsible for making high Mach number shocks
effective sites for electron acceleration.
\end{abstract}





\section{Introduction}\label{sect:introduction}

Collisionless shocks are a key component of many astrophysical
systems, since they act as sites of particle acceleration and
thermalization, converting upstream bulk flow energy into 
downstream thermal energy, and at the same time transferring
energy to a small fraction of energetic particles. The internal
structure of a collisionless shock determines the thermalization
process, and a controlling factor is the magnetic geometry
of the shock, usually described by the angle $\theta_{Bn}$
between the shock normal and the upstream magnetic field direction.
Quasi-parallel ($\theta_{Bn}<45^\circ$) and quasi-perpendicular
($\theta_{Bn}>45^\circ$) shocks have very different internal
structure, different thermalization processes, and are also
predominantly associated with different acceleration mechanisms:
diffusive (Fermi) acceleration and shock drift acceleration (SDA),
respectively.


For quasi-perpendicular shocks in the heliosphere, the internal
structure is well established, both observationally and with
simulations \citep[e.g.,][]{burgess1995}. At sufficiently high Mach
numbers (supercritical) ion reflection is required to provide
the required downstream heating. Specularly reflected ions
gyrate in front of the shock (due to the quasi-perpendicular
geometry) forming a ``foot'' in the magnetic field and density
ahead of the main ramp. The reflected ions, as they transit
downstream, are also responsible for an overshoot-undershoot
structure immediately after the ramp. The overshoot means that
the magnetic field magnitude peaks at a larger value than
the simple expectation from the shock conservation relations
for a high Mach number perpendicular shock. At subcritical
Mach numbers, there is little or no ion reflection, and
dispersion and/or dissipation provides the limiting
process against nonlinear steepening.

Energetic electrons up to 100 keV are commonly observed at interplanetary
shocks and at planetary bow shocks. They are also invoked in
mechanisms of radio emission in the solar corona and the interplanetary
medium.
\citet{tsurutani+lin1985}
reported observations of interplanetary shocks with
associated spikes and step-like increases in
electron fluxes in the energy range $2 - 20$ keV.
The largest events were at shocks with 
$\theta_{Bn} \stackrel{<}{\sim} 70^\circ$ and
with high (relative to the data set) shock Mach number. 
Some shock crossings had no accelerated electron signature, and these
were those with low Mach number and small $\theta_{Bn}$.
Higher energy electrons, up to at least 100~keV
can sometimes be associated with interplanetary
shocks \citep[e.g.,][]{simnett++2005}.

\citet{anderson++1979} observed electrons up to 100 keV ahead
of the Earth's bow shock, in a thin sheet just downstream of the
tangent surface formed from the solar wind field lines touching
the curved bow shock surface. It was found that the most
energetic electrons ( $>5.3$ keV) originated on the bow shock where
$\theta_{Bn}>85.5^\circ$, i.e., closest to the magnetic tangent
point. It was clear that the most energetic electrons were
associated with the quasi-perpendicular shock, indeed the
near-perpendicular shock.
These observations led to the suggestion that reflection of
incident electrons was important to the acceleration process.
\citet{holman+pesses1983} outlined a mechanism for type II solar
radio bursts in which upstream electrons were reflected and
energized by shock drift acceleration (SDA).

Based on the idea of reflection at
the quasi-perpendicular shock
by conservation of magnetic moment, \citet{wu1984}
and \cite{leroy+mangeney1984} developed an analytic model for
electron acceleration out of the thermal distribution. The theory
was developed for a planar, time-steady shock with scatter-free
propagation for the electrons. The de Hoffman-Teller frame
(HTF) is the shock frame in which the upstream flow is directed
along the magnetic field so there is zero motional electric field.
In the HTF
particles conserve their energy and magnetic moment if their
propagation is scatter-free and the scale lengths and time scales
are long compared to their cyclotron motion. Reflection
is described by magnetic mirroring (or adiabatic reflection),
with some additional effects at low
energy due to the cross-shock potential. In the normal incidence
frame (NIF) the particles drift in the gradient of the magnetic
field along the motional electric field direction, and thus gain
energy. \cite{kraussvarban+wu1989} have shown the equivalence of the HTF and
NIF descriptions of the energization process.
Since scatter-free propagation is assumed,
the reflected distribution function and moments can be calculated
by Liouville mapping from the incident distribution.
An electron which is reflected has to satisfy the conditions that
its magnetic moment is appropriate for magnetic mirroring, and that
its parallel velocity after reflection is high enough to escape
the shock (i.e., in the HTF its parallel velocity is directed upstream).
As $\theta_{Bn}$ approaches 90$^\circ$ the frame transformation
velocity from NIF to HTF increases, and the portion of the
incident distribution that will mirror, and have an upstream
directed 
velocity in the HTF, has an increasingly high energy in the
incident flow frame. Thus, the reflected accelerated density
depends strongly on the form of the incident distribution
above thermal energies. This produces a paradox: acceleration
due to magnetic mirroring will give the highest energies
for shocks closest to perpendicular, but the same shocks produce
the lowest reflected fraction. This situation is ameliorated
by using either a core plus halo or kappa distribution upstream, as is
typical for solar wind electrons
\citep[e.g.,][]{fitzenreiter++1990}.

The results of adiabatic reflection theory have been tested,
and broadly confirmed for energies up to at least $10 - 20$ keV,
using
a combination of test particle integration and self-consistent,
one-dimensional hybrid simulation of the shock fields
\citep{kraussvarban++1989}.
The use of self-consistent shock simulations removes the need
to make model assumptions about the magnetic field and electric
potential profiles. It was also realized that, in the
scatter free approximation, the electrons could travel a long
distance along the magnetic field direction during their
reflection process, and thus shock curvature would become important.
Modeling of a curved shock was carried
out which revealed a flux focusing effect, so that particles
entering at $\theta_{Bn}$ close to $90^\circ$ drift and exit
at a slightly lower value \citep{kraussvarban+burgess1991}.
This latter result is again for the case of scatter-free motion.

The effects of a curved shock, with finite width, were also investigated
by \citet{vandas1995b} using a model shock profile. This work
was extended to include comparisons between modeled
and observed upstream anisotropies and energy
spectra at the Earth's bow shock \citep{vandas2001}.
The assumptions of magnetic
moment conserving reflection and scatter-free Liouville mapping
make it possible to construct the distribution function of
reflected electrons throughout the foreshock, i.e., the region ahead
of, and magnetically connected to the bow shock \citep{cairns1987}.
This has been used for two purposes: Firstly, in order to make
comparisons with observations of low energy electrons in the
terrestrial foreshock \citep{fitzenreiter++1990},
and, secondly, as a component in models of foreshock radio emission.
For example,
\citet{kuncic++2004} describe a quantitative model of fundamental
and second harmonic emission in the Earth's foreshock, using
electron beam properties derived from adiabatic reflection of
electrons out of the solar wind distribution. \citet{knock++2003}
constructed a similar model to explain interplanetary Type II radio bursts,
based on acceleration at a large scale ``ripple'' on the surface
of an interplanetary shock. Electron acceleration in a ``wavy'' shock
front was also modeled by \citet{vandas+karlicky2000} to explain
solar Type II radio bursts.

Adiabiatic reflection theory has several positive aspects: It produces
qualitative agreement with observations in terms of the origin and
energies of accelerated electrons. Furthermore, observations
at lower energies ($<1$ keV) often show loss-cone features in
agreement with magnetic mirroring \citep[e.g.,][]{fitzenreiter++1990}.
However, for the more energetic electrons, the theory fails to explain the
observed energy spectra and anisotropies at the shock \citep{vandas2001}.
Observations at the Earth's bow shock in the energy range $1-20$ keV
show that the suprathermal electron flux appears as an inverse power
law (with exponent $3-4$ for the phase space density)
which emerges smoothly from
the downstream thermal distribution, with fluxes peaking
just downstream, and the highest energies ($>5$ keV) appearing at
the shock overshoot where the distribution
is nearly isotropic \citep{gosling++1989}. The backstreaming (so-called
reflected) component is seen in the shock ramp, and this was
interpreted as escaping particles which have been energized at,
and just downstream of the shock.

The observation of a high energy power law tail for the
shock accelerated electrons, both upstream and downstream, and
small downstream anisotropy, are difficult to explain
in the context of adiabatic reflection, which, as in SDA, is a
single encounter process. \citet{vandas2001} concluded that some other
process, such as pitch angle scattering, must modify the reflection
process, in order to explain observations. Modeling of the effects
of such scattering in a curved shock was carried out by
\citet{kraussvarban1994}, although the scattering was introduced
in an ad-hoc manner. 

So far we have concentrated on the idea that the acceleration is
due to the shock acting as a moving magnetic mirror. However, electron
acceleration is also possible by a high level of turbulence within
the shock layer. For example, \citet{cargill+papadopoulos1988}
investigated the possibility that a Bunemann instability of the
reflected ion beam at high Mach numbers can lead to strong electron
heating. Electron heating and (via formation of nonthermal tails)
acceleration have been studied with full particle (kinetic ion and
electron) plasma shock simulations
\citep[e.g.][]{bessho+ohsawa2002,lembege+savoini2002}.
The formation of an electron foreshock at a curved shock has
been studied with a full particle simulation \citep{savoini+lembege2001}.
\citet{schmitz++2002} presented full particle
simulations where accelerated electrons interacted with waves in the
foot of the shock, and  could be trapped in electron phase
space hole structures, thereafter they interacted with the shock 
in a manner approximately conserving magnetic moment. Although such
full particle PIC simulations give an insight into the various
physical interactions, they are hampered by computational
constraints which mean the use of non-realistic ion-electron mass 
and ion-electron plasma frequency ratios. It has been shown that using
a low value for the mass ratio in such simulations can lead to
behavior which is not seen when the correct value is used
\citep{scholer++2003}. Thus, there are some questions about the
applicability of the results of such simulations to interplanetary
and planetary shocks; their results may be more applicable to high
Mach number relativistic shocks.

The work in this paper starts from the idea that the quasi-perpendicular
shock acts as a magnetic mirror, and that, for initial energies just above 
thermal, electron scale fluctuations can be neglected, so that the
hybrid simulation method is appropriate
\citep{kraussvarban++1989}. It is shown, using
a combination of self-consistent and test particle simulations,
that the inclusion
of structure along the magnetic field lines (and across the surface
of the shock) produces scattering within the shock layer which 
fundamentally changes the resulting upstream and downstream electron
distributions. In particular, we find that power law energy spectra can be
produced both downstream as well as upstream, and that the range
of $\theta_{Bn}$ over which acceleration is effective broadens, so the
overall efficiency of the process increases.

The paper is organized as follows: Section \ref{sect:simulations}
describes both the hybrid and test particle parts of the simulations;
Section \ref{sect:structureinshock} demonstrates the types of
shock structure that are found in the hybrid simulations;
Section \ref{sect:electronenergyspectra} gives the resulting
energetic electron spectra; finally, Section \ref{sect:summary}
summarizes and briefly discusses the results.

\section{Simulations} \label{sect:simulations}

The method we use is similar to earlier work
\citep{kraussvarban++1989}.
The major difference is that
a spatially two dimensional hybrid simulation is used, rather
than the one dimensional simulation used in earlier work.
Two dimensional hybrid simulations of the quasi-perpendicular
shock were first carried out by \cite{winske+quest1988}, who
showed the presence of large amplitude fluctuations in the
field and density at the shock transition. These fluctuations
of the surface of the shock have been analyzed in terms of
turbulent ripples \citep{lowe+burgess2003}.

The simulation is carried out in two stages. First, the electric
and magnetic fields for the shock transition
are generated using a hybrid plasma simulation.
Data for all time steps and grid points is stored.
The hybrid plasma simulation models the electron response as
a fluid. Ions are modeled using simulation macroparticles as
in the standard PIC method. The hybrid method has a number
of advantages suited to the collisionless shock problem.
Ion kinetic effects are retained, including ion reflection which is
crucial to correct modeling of the shock structure. On the other
hand, the fluid electron response means that relatively long
times and large spatial domains can be simulated. This is important
for the interaction of electrons with the shock, since, at the
energies and Mach number considered here, they convect into the
shock at the bulk flow speed.
At a supercritical shock the convection time
of a field line through the shock structure is of the order of the
ion cyclotron time, since the foot structure scales with the Larmor
radius of the reflected ions. Thus to study properly the process of electron
energization the shock fields must be simulated on ion time scales.
The electron fluid response is also appropriate to planetary bow shocks
and interplanetary shocks, where strong electron heating is only
rarely observed.

The second stage of the simulation consists of integrating the 
equations of motion of an ensemble of test
particle electrons in the time varying fields
obtained from the hybrid simulation. Careful interpolation of
the fields in time and space is required to ensure accuracy,
otherwise artificial scattering of the test particles may be
introduced. The initial conditions used in this work are
to follow a shell of mono-energetic electrons released upstream,
until they reach either specified upstream or downstream collection
points.

We now give details of the simulation.
The hybrid simulation models the plasma with macro-particle ions
(protons) and a massless, charge neutralizing electron fluid with
an adiabatic equation of state. The simulation method is described in
\cite{matthews1994}.
Distances are normalized to the ion inertial length $c/\omega_{pi}$,
time to the inverse cyclotron frequency $\Omega_{ci}^{-1}$, velocity
to the Alfv\'en speed $v_A$; all normalizations use upstream values.
The magnetic field is normalized to its upstream value, $B_0$.

The shock is launched by the injection method, whereby plasma flowing
in the $x$ direction is introduced at the left hand boundary of
the simulation. The right hand boundary is treated as a perfectly 
reflecting wall. The simulation is periodic in the $y$ direction.
The shock forms by reflection of the flowing plasma with the wall,
and propagates against the flow (in the simulation frame) in the
$-x$ direction. Thus, the simulation frame is a normal incidence
frame, with the flow-normal angle $\theta_{Vn}=0.$
The plasma inflow speed $M_i$ (in units of $v_A$) is specified,
and by the shock motion in the simulation frame its 
Alf\'en Mach number $M_A$ can be found.
The simulation domain is 100 by 20 
$c/\omega_{pi}$, with a cell size of
0.2 by 0.2 $c/\omega_{pi}$; the timestep is 0.01 $\Omega_{ci}^{-1}$.
A resistivity is included with a small, ad hoc value to suppress
very short scale gradients.
In order to model the reflected
gyrating ions accurately, and to reduce statistical noise, which is
important to reduce unrealistic scattering of the test particle electrons,
a fairly large value of 200 ions per cell (upstream) is used.

The equations of motion for the test particle electrons are
integrated with a fourth order scheme with appropriate conservation
properties \citep{kraussvarban++1989,thomson1968}. Non-relativistic
motion is assumed, appropriate to the energy range studied
here. The size of
time step used is varied adaptively in order that the local error,
for both position and velocity, achieves some given bound. The local
error is estimated from the residual change from reducing the
step size. We have found that the use of an adaptive step size
is essential for accurate and efficient following of the test
particles, especially the most energetic particles in our
simulations. The step size was allowed to vary between the bounds
of 0.1 and 2 10$^{-5}$ $\Omega_{ce}^{-1}$. It was verified that the
results did not depend on the choice of lower limit, and that
the target error bound was small enough for the results not to
change, in a statistical sense, when the error bound was reduced.

The time step used for the electron integrator is very much smaller
than that from the hybrid shock simulation, and thus
it is essential to use a smooth interpolation scheme
for the values of the fields at the test particle position. We use a
bicubic scheme for spatial interpolation and a piece-wise linear
interpolation in time. Because of the high parallel speed of the electrons,
they can travel much further than one grid cell in a hybrid time step,
so smooth spatial interpolation is essential when following the
electron motion on the cyclotron time scale. For example, using 
only linear interpolation in space introduces discontinuities
in slope which adds a noise component to the fields experienced
by the test particle electrons as they cross the hybrid simulation
cell boundaries. This can lead to the particles undergoing
effective but artificial and unrealistic pitch angle scattering
and hence acceleration.

In this paper we follow ensembles of initially mono-energetic
electrons, energy $E_i$,
released at a fixed distance upstream of the shock, uniformly
distributed over the $y$ extent of the hybrid simulation. The
velocity space distribution is a sphere centred on the upstream
flow velocity.
Because of the high parallel speeds of some electrons, they can
outrun the shock from the point of release and never interact with
the shock. Based on the upstream directed parallel speed and an
average shock speed in the simulation frame, particles which will
definitely not interact with the shock are excluded from the
simulation, although included for the purpose of normalization of
the energy spectrum. This increases the efficiency of the simulation
by improving the statistics of particles which interact with 
the shock, thereby increasing the dynamic range of the spectra.
The parameter for deciding whether a particle will interact with the
shock is chosen conservatively, so that in the spectra shown some
particles will be included which have not interacted with the shock,
even though the majority are correctly excluded. The effect
is that the flux level at the injection energy excludes the
majority, but not all, of the non-interacting particles.

The test particles are released after sufficient time
so that the shock and its structure and downstream region
has become developed; for the simulations shown here this at
T=10 $\Omega_{ci}^{-1}$ in the hybrid simulation.
In order to convert from energy in keV to normalized
units in the hybrid system, a value for the ratio $v_A/c$ of
5000 is used, which corresponds to an Alfv\'en speed of 60 km s$^{-1}$,
typical of the solar wind. The test particles are then followed as
they interact with the shock, until they reach fixed positions
relative to the shock either upstream or downstream. In this work
the upstream release distance is +5 $c/\omega_{pi}$, and the
upstream and downstream collection distances are +7
and $-4$ $c/\omega_{pi}$, respectively. The downstream collection
point is chosen as the location where the $y$-averaged magnetic
field magnitude first reaches its average downstream value after the
shock overshoot. All these distances are relative to an
instantaneous shock position defined as the location where the
$y$-averaged magnetic field first passes the value $2B_0$.
Finally, in this paper the results of electron interaction
with the shock are shown as the energy spectra of the differential
energy flux $dJ/dE\sim Ef(E)$ where $f(E)$ is the omnidirectional
particle energy number distribution.

\section{Results} \label{sect:results}

\subsection{Structure in Shock Transition} \label{sect:structureinshock}

\begin{figure}
\epsscale{.80}
\plotone{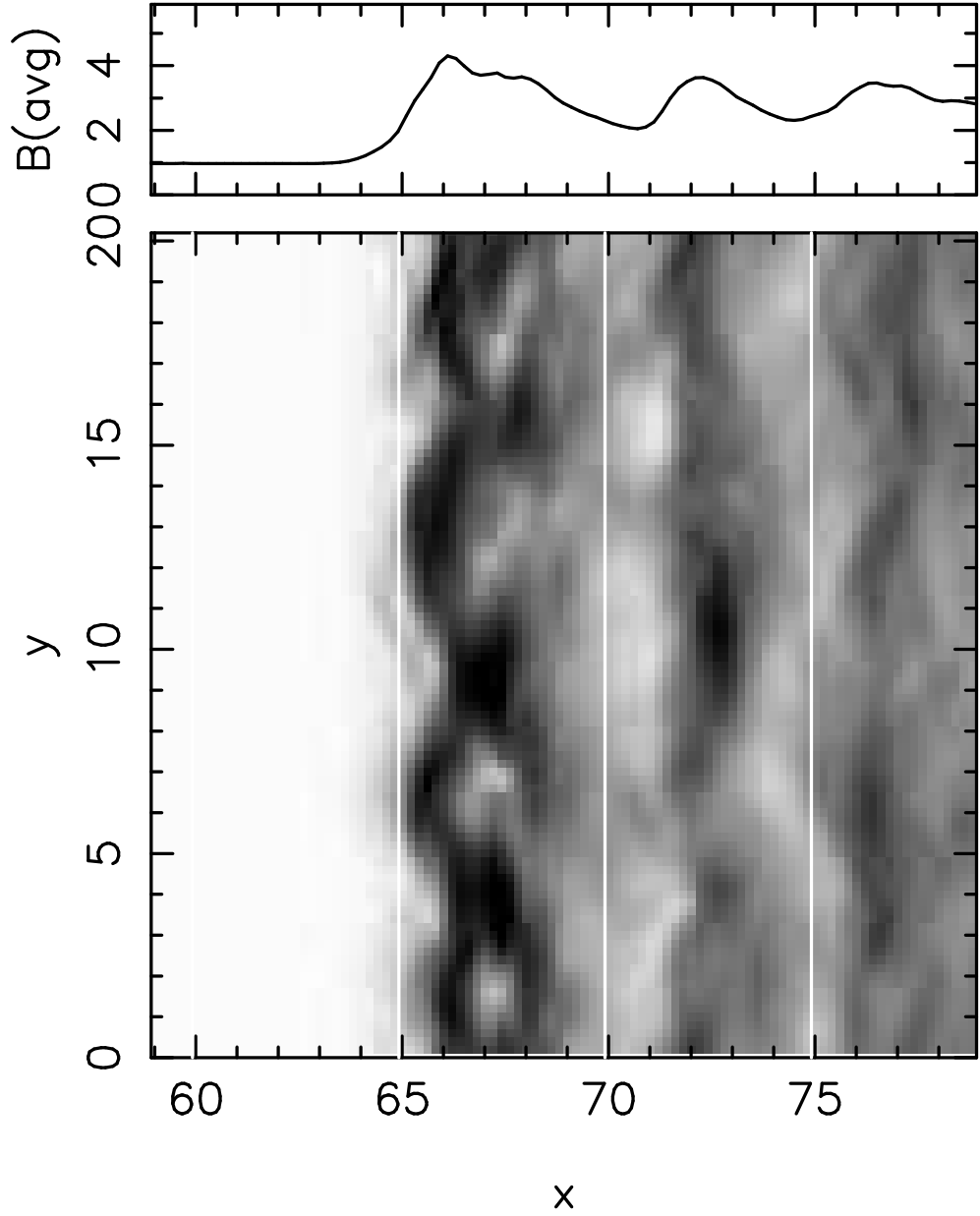}
\caption{Gray scale map of magnetic field magnitude 
for Case A (see text) at time $T=20 \Omega_{ci}^{-1}$ in
the hybrid simulation. The white-black scale is $0.9 - 5.0$.
The magnetic field averaged over the
$y$ direction is shown in the upper panel.
Only a small region of simulation around the shock is shown.
\label{fig:th87bp5m3p5}}
\end{figure}

In this section we will illustrate the shock structure seen in
hybrid simulations for a range of
different shock and simulation parameters that will be used for the
electron test particle simulations. The first example (Case A, Figure
\ref{fig:th87bp5m3p5}), and
reference case, is for $\theta_{Bn}=87^\circ$, inflow velocity
of $M_i=3.5$, and resultant shock Alfv\'en Mach number of $M_A=5.1$.
In all examples
here the ion and electron upstream plasma beta is $\beta_i=\beta_e=0.5$.
Figure \ref{fig:th87bp5m3p5}, upper panel, shows the magnetic field magnitude
$B$ averaged over the $y$ direction. This average is rather stable over time,
with only slight changes in form. The presence of the usual components
of super-critical quasi-perpendicular shock structure is evident: foot, ramp,
overshoot, and undershoot.
Using the definition of the 
shock position given above, an instantaneous shock speed can be 
calculated. This shows some quasi-periodic variations on both short
$\sim$0.1 $\Omega_{ci}^{-1}$ and longer $\sim$1.2 $\Omega_{ci}^{-1}$
time scales.
This variation indicates small changes in the gradient and maximum
value of the
average $B$ profile; such behavior is also seen in one dimensional
hybrid simulations.

Figure \ref{fig:th87bp5m3p5} (lower panel) shows a gray-scale map of the
magnetic field magnitude in a small range of $x$ around the shock transition.
The most obvious feature is that there is considerable structuring
along the shock surface. Locally the field gradients are considerably
larger than for the average profile. The shock surface has a rippled
appearance, so that the local gradient normal changes in direction along
the surface. The figure just shows the shock surface at one time, but
the structure is highly dynamic on time scales of between 0.1 and 1.0
$\Omega_{ci}^{-1}$, with the pattern of ripples moving and changing in
form. This
can lead to rapid changes in the local angle between the upstream
magnetic field and the gradient normal. Figure \ref{fig:th87bp5m3p5}
shows the shock at a time when the rippling appears quasi-sinusoidal,
but this feature is not constant; at other times the rippling is less
coherent. The dynamic
behavior of the shock structure has been analysed by
\cite{lowe+burgess2003} in terms of ripples propagating across the
shock at approximately the Alfv\'en speed at the shock overshoot.
Although we only show its magnitude, the perturbations in the shock
layer occur strongly in all components of the magnetic field and 
the density \citep{winske+quest1988}.

All the cited work on two-dimensional hybrid shock simulations,
and the above example, are for the case where the upstream magnetic
field direction lies in the $x-y$ simulation plane. This allows
the existence of perturbations with wave vectors parallel to the
magnetic field direction, e.g., Alfv\'en ion cyclotron waves, which
are known observationally to exist downstream of the supercritical
quasi-perpendicular shock, and which therefore play an important role in
ion thermalization at the shock transition \citep{mckean++1995}.
However, it is possible to place the upstream magnetic field direction
in the $z-x$ plane, i.e., out of the simulation plane,
so that the simulation cannot contain any perturbations
propagating parallel to the field and transverse to the shock normal.
This has the effect of suppressing Alfv\'en ion cyclotron and other
predominantly parallel propagating waves.

\begin{figure}
\epsscale{.80}
\plotone{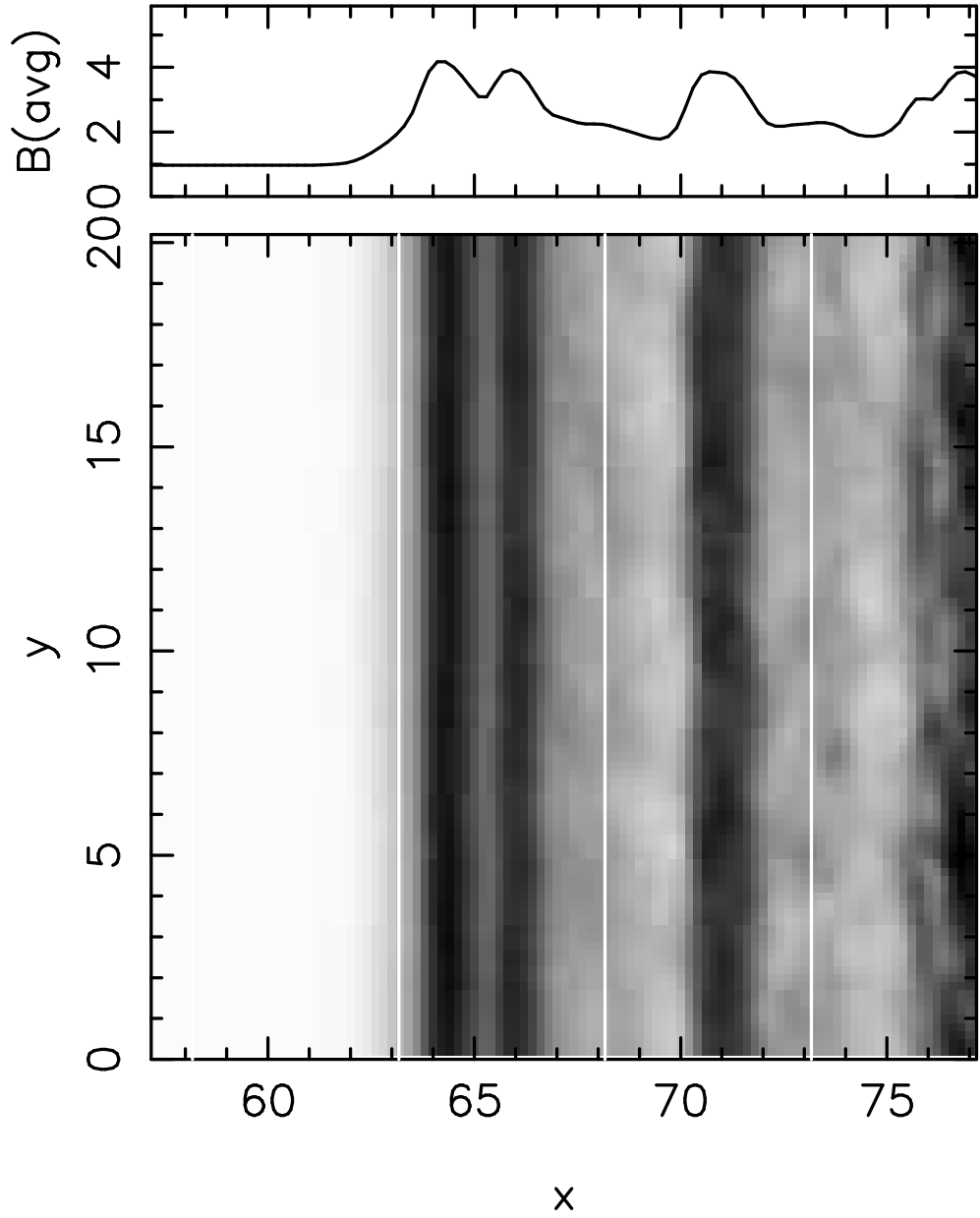}
\caption{As Figure \ref{fig:th87bp5m3p5}, but for Case B (see text).
\label{fig:th87bp5m3p2oop}}
\end{figure}

Figure \ref{fig:th87bp5m3p2oop} illustrates this case (Case B), with
$\mathbf B$ out of the simulation plane for the
parameters $\theta_{Bn}=87^\circ$, inflow velocity
of $M_i=3.2$, and resultant shock Alfv\'en Mach number of $M_A=5.1$.
The inflow velocity has been adjusted so that the shock Mach
number is the same as for Case A. This adjustment is necessary
because the difference in the shock structure, which is clearly
evident, leads to a different downstream thermalization, changing
the propagation speed of the shock in the simulation frame.
Although the appearance of the average profile
is very similar to Case A, it is clear that putting the
upstream field out of the simulation
frame strongly suppresses any structuring of the shock transverse
to the normal; the shock structure effectively becomes
one dimensional. We should note that two-dimensional
structuring of the shock can be found in simulations with $\mathbf B$
out of the simulation plane at higher Mach numbers.
Although Case B is unrealistic, because
perturbations propagating parallel to $\mathbf B$ are suppressed,
we will be using it to demonstrate the role of dynamic shock structure
in electron acceleration. 

\begin{figure}
\epsscale{.80}
\plotone{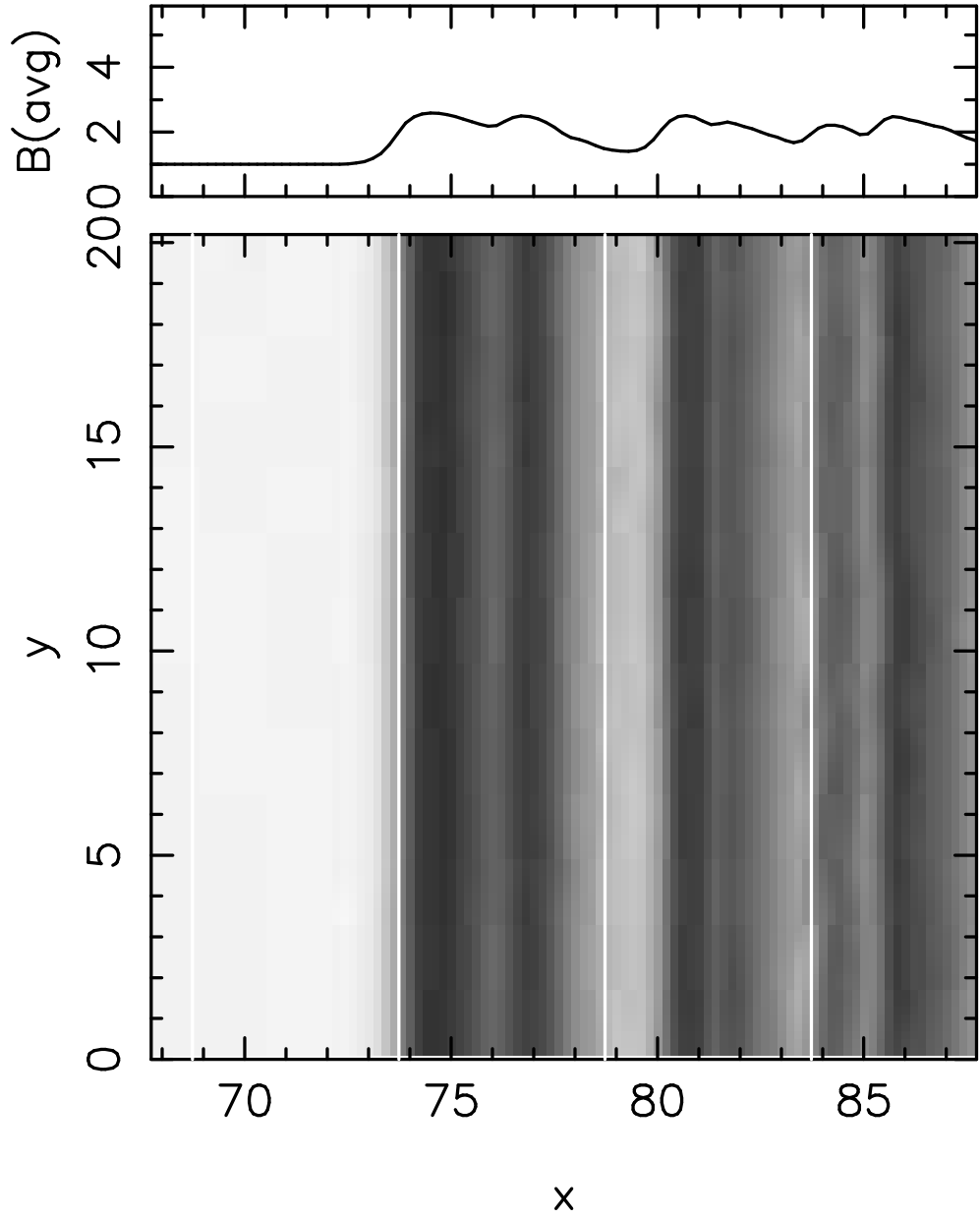}
\caption{As Figure \ref{fig:th87bp5m3p5}, but for Case C (see text).
The white-black scale is $0.9 - 3.0$.
\label{fig:th87bp5m1p5}}
\end{figure}

\cite{lowe+burgess2003} reported that the rippled structuring of
the shock is associated with the presence of an overshoot in the
average magnetic field profile, i.e., the shock has to have a
high enough Mach number to be supercritical, with the presence
of reflected gyrating ions at the shock transition. We provide
an example of this in Figure \ref{fig:th87bp5m1p5} (Case C), which
shows the shock transition for a low Mach number shock with
$\theta_{Bn}=87^\circ$, inflow velocity
of $M_i=1.5$, and resultant shock Alfv\'en Mach number of $M_A=2.1$.
The upstream magnetic field direction is in the simulation plane,
as for Case A, but in comparison there is no appreciable overshoot
in the magnetic field profile, and any rippling or variations of
the shock surface is at a very low level. Again the shock
structure is virtually one dimensional.

\begin{figure}
\epsscale{.80}
\plotone{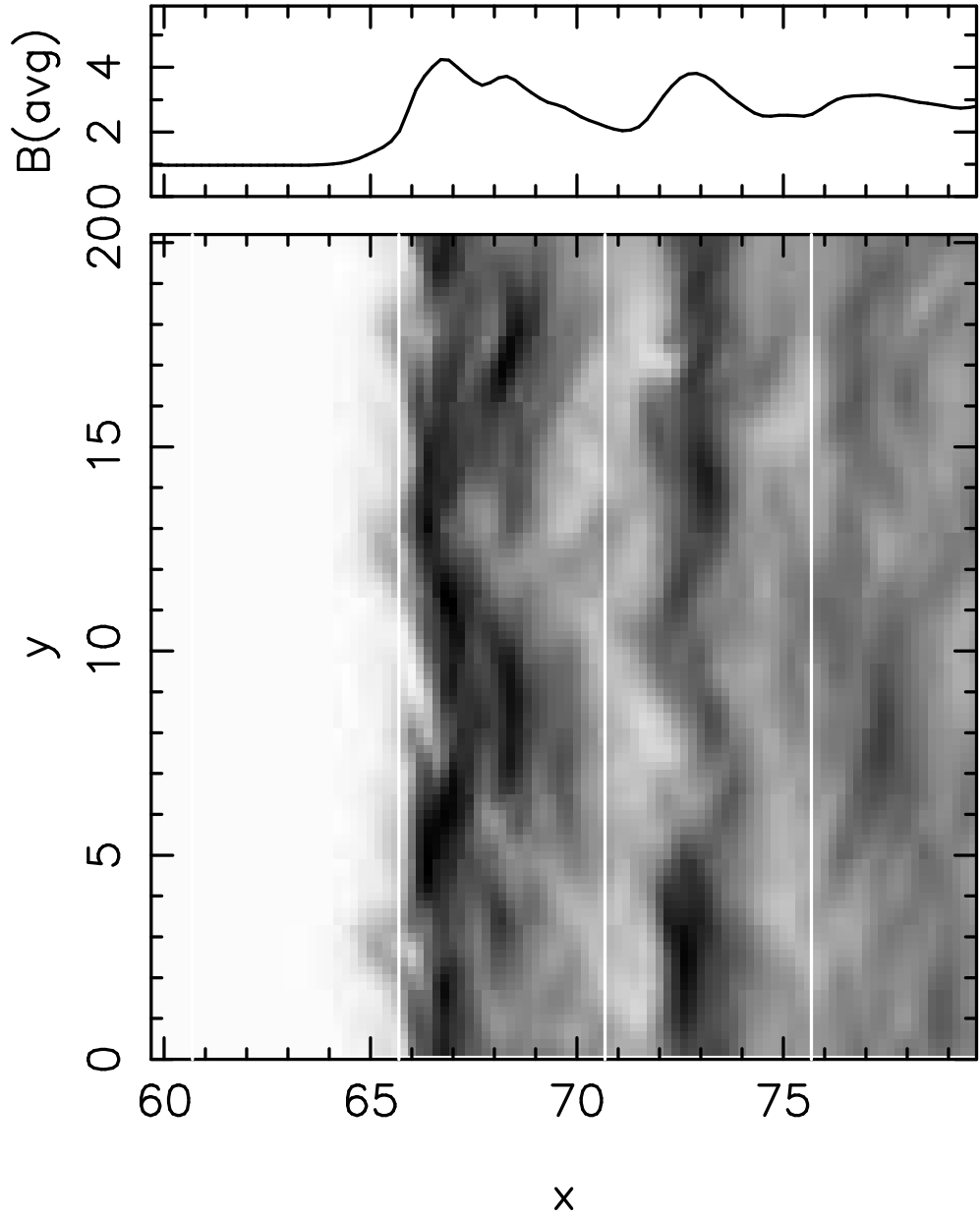}
\caption{As Figure \ref{fig:th87bp5m3p5}, but for Case D (see text).
\label{fig:th80bp5m3p5}}
\end{figure}

So far a shock normal angle of
$\theta_{Bn}=87^\circ$ has been used for all the examples.
We will be showing
the effect of $\theta_{Bn}$ on electron acceleration, and Figure
\ref{fig:th80bp5m3p5} (Case D) shows the shock structure for
the lower value of $\theta_{Bn}=80^\circ$, but with other
parameters the same as Case A. Comparing with Case A, the profile
of the average magnetic field is very similar, and the two dimensional
structuring is also basically similar. There are some differences
of appearance; the shock transition for the lower $\theta_{Bn}$ case has
a slightly more fractured appearance, with the more evident
presence of precursor waves, presumably associated with oblique whistlers.
Again, as for Case A, the shock transition structuring is highly
dynamic in time and space.

\subsection{Electron Energy Spectra} \label{sect:electronenergyspectra}

\begin{figure}
\epsscale{.80}
\plotone{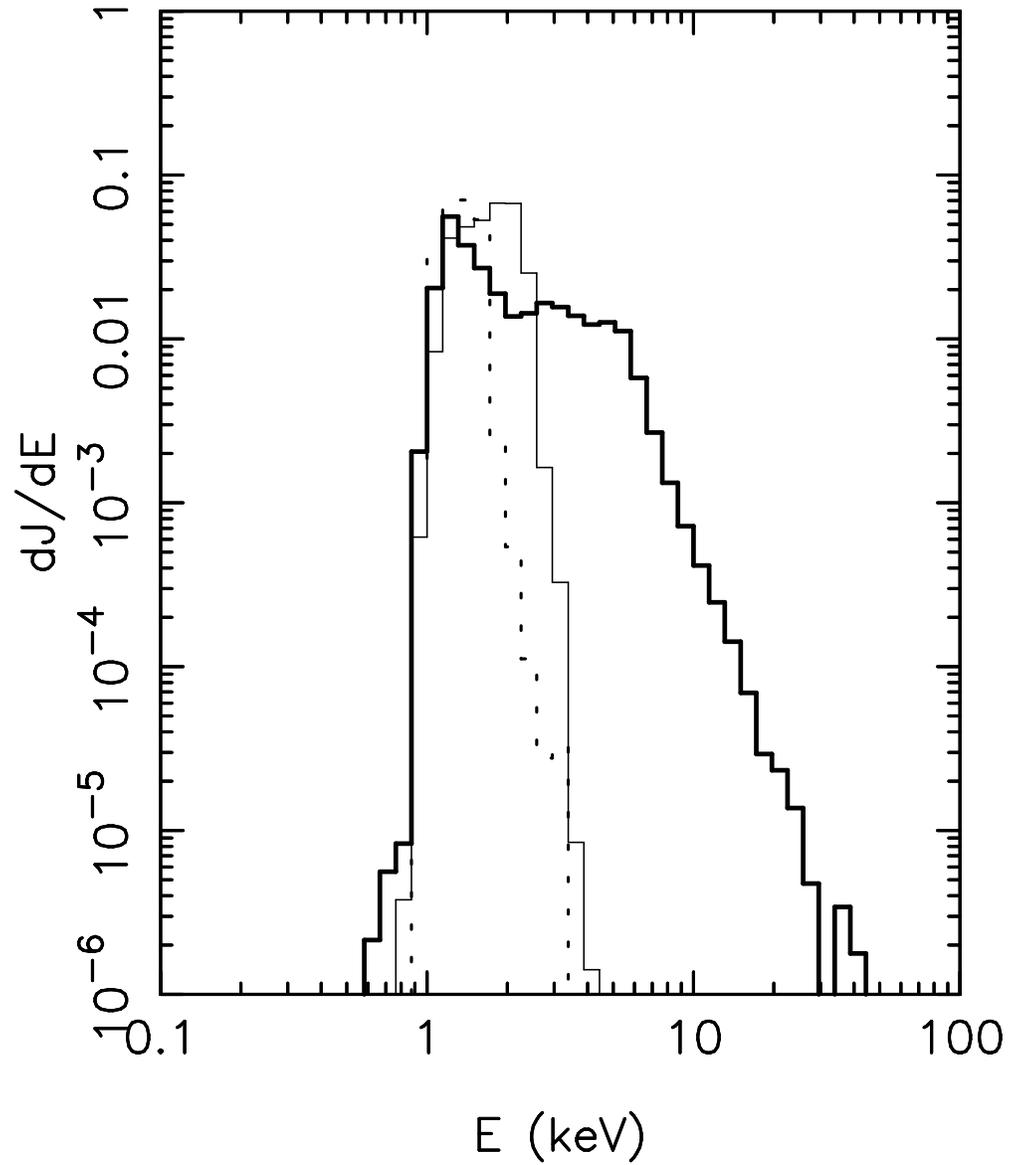}
\caption{Upstream differential flux energy spectra for an injection
energy of 1~keV for: Case A (thick line), Case B (thin
line), and Case C (dotted line). Case A has a shock surface with dynamic
rippled structure, while the others have a quasi-one dimensional structure.
\label{fig:th87:1keV:cf}}
\end{figure}

The structure and variability within the shock transition
modeled by the two dimensional
hybrid simulation has a major effect on the spectra of shock
accelerated electrons in the energy range up to about 10~keV.
This is demonstrated in Figure \ref{fig:th87:1keV:cf} which 
shows the upstream energy spectra for an initial energy $E_i=1$~keV
for Case A (reference simulation), Case B ($\mathbf B$ out
of simulation plane) and Case C ($\mathbf B$ in simulation plane,
but low $M_A$). Case A shows considerable structuring and
variability, whereas Cases B and C show little or no structuring
of the shock front. Three important features are evident in
the spectrum for Case A: The maximum energy achieved 
is much greater than for the other cases, by a factor
of almost 10. There is a region, roughly $2 - 6$ keV,
where the variation of the differential energy flux is almost flat.
Finally, above 6 keV the spectrum falls off with a power law
dependence with a slope of approximately $-4.6$.

The cases without a rippled shock structure
have much smaller maximum energies, and a sharp cut-off
with increasing energy. The peak energies for these cases are
in line with the prediction of adiabatic reflection theory, as
is the sharp upper limit. The low Mach number spectrum (Case C) shows
a smaller peak energy, again consistent with adiabatic reflection
theory. For case B, which has the same Mach number as Case A, the
flux at low energies is greater than for Case A, but then Case A
displays considerably higher (by many orders of magnitude) fluxes
at higher energy.

\begin{figure}
\epsscale{.80}
\plotone{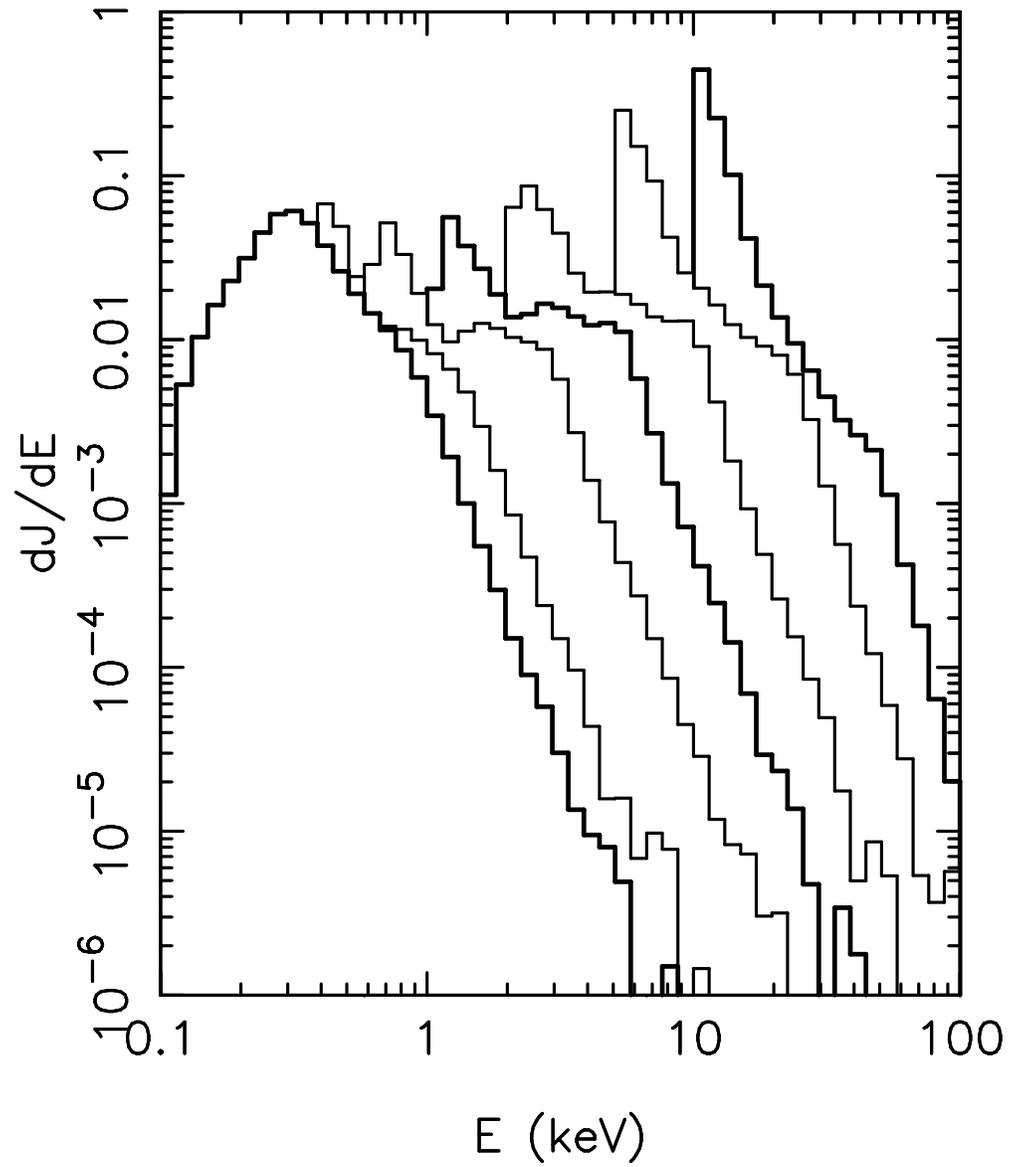}
\caption{Energy spectra for Case A for electron injection
energies of 100~eV, 200~eV, 500~eV, 1~keV, 2~keV, 5~keV, and 10~keV.
\label{fig:th87bp5m3p5:cf:100:10keV}}
\end{figure}

Figure \ref{fig:th87bp5m3p5:cf:100:10keV} shows the spectra
for a range of injection energies $E_i$ from 100 eV to 10 keV
for case A, which has dynamic
structuring across the shock surface and along the magnetic
field lines. All spectra show a transition to a power law fall off
above a certain energy. The spectrum for $E_i=100$~eV shows, after
its peak value, a fall off with a change of slope around 1 keV.
For $E_i>100$~eV all spectra have a peak in the range
up to about $2E_i$, typical of a single reflection interaction.
Recall that the algorithm for removal of particles which
are headed away from the shock introduces some error at the lower
edge of the spectra.

\begin{figure}
\epsscale{.80}
\plotone{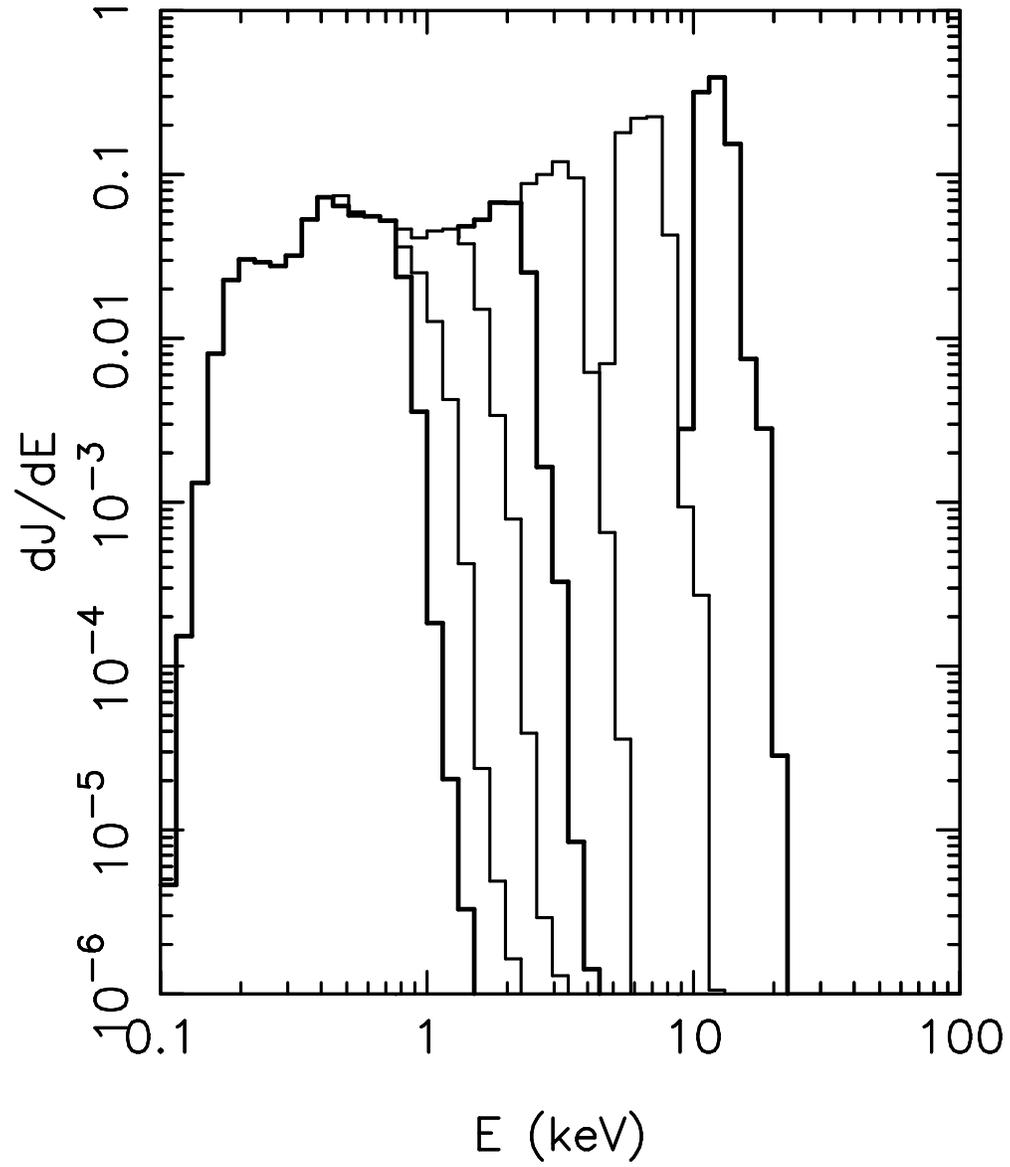}
\caption{As for Figure \ref{fig:th87bp5m3p5:cf:100:10keV}, but for
the shock Case B which does not show a structured shock front.
\label{fig:th87bp5m3p2oop:cf:100:10keV}}
\end{figure}

The most prominent feature is that part of the spectrum which is almost
flat for injection energies in the range 500 eV to 2 keV. This
feature begins to be evident around $E_i=200$~eV, for which there
is a narrow peak, a small region of shallower slope, and then
a steeper power law fall off. Beyond $E_i=2$~keV, the spectrum
does not have a flat plateau region, but instead a region of shallower
slope, and the slope steepens with increasing injection energy.
Thus, by $E_i=10$~keV the spectrum shows a peak  up to 20 keV, then a fall
off which changes slopes and steepens at around 50 keV.
In contrast to the results for Case A,
Figure \ref{fig:th87bp5m3p2oop:cf:100:10keV} shows the
spectra for a similar range of injection energies, but for
Case B, where the structuring along the shock front and magnetic
field direction has been suppressed by putting the 
upstream magnetic field direction out of the simulation plane.
In this case all spectra
show an extremely rapid fall off with increasing energy, consistent
with reflection from a one dimensional shock structure.

\begin{figure}
\epsscale{.80}
\plotone{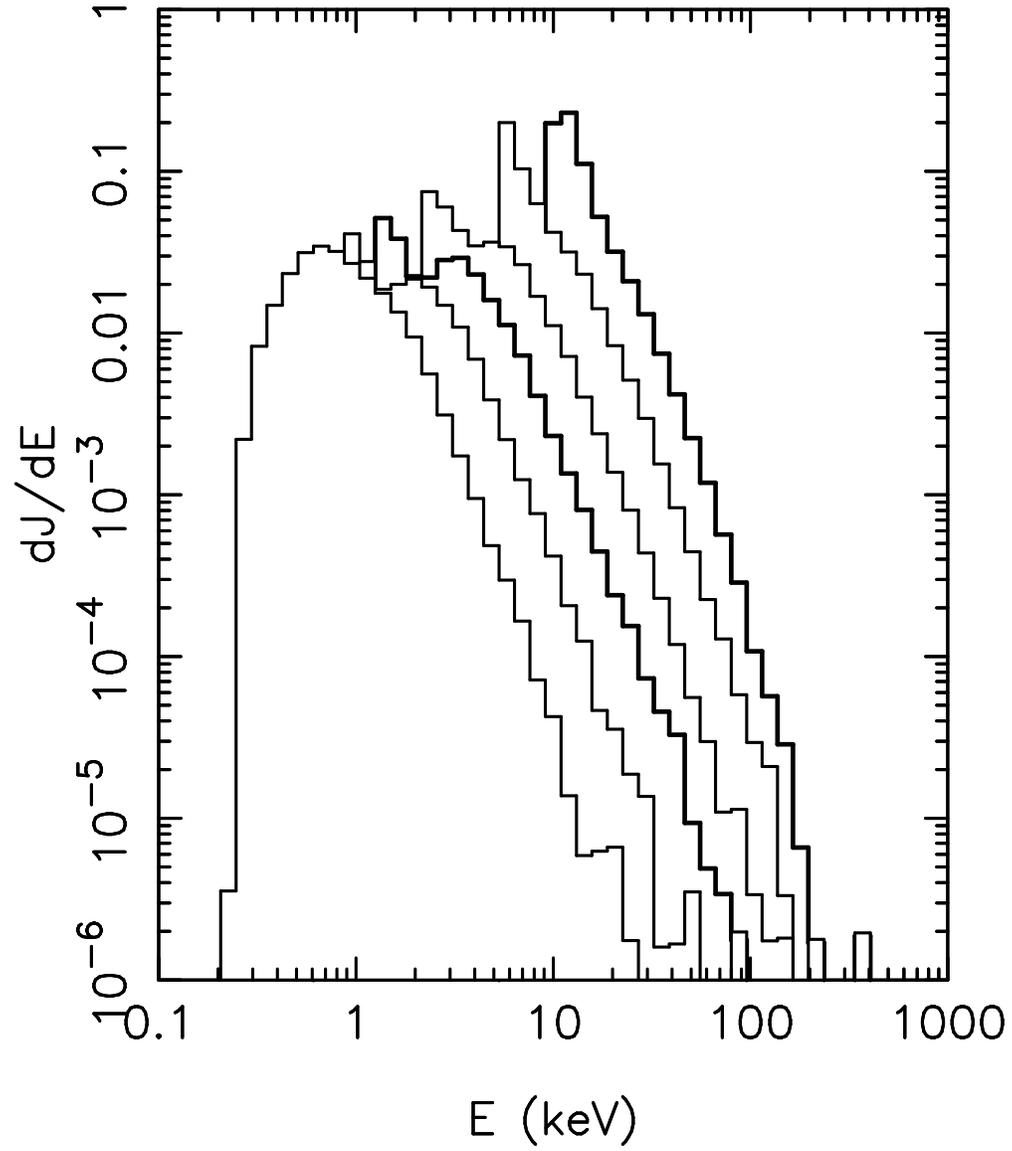}
\caption{Energy spectra for a shock similar to Case A, but with
$\theta_{Bn}=88^\circ$, for electron injection
energies of 200~eV, 500~eV, 1~keV, 2~keV, 5~keV, and 10~keV
\label{fig:th88bp5m3p5:cf:200:10keV}}
\end{figure}

The shock normal angle $\theta_{Bn}$ is a crucial parameter in
adiabatic reflection theory for electron acceleration. Figure
\ref{fig:th88bp5m3p5:cf:200:10keV} shows the spectra
for $E_i$ from 200~eV to 10~keV for a simulation with the
same parameters as Case A, but with $\theta_{Bn}=88^\circ$.
The overall behavior is very similar to the results
for $\theta_{Bn}=87^\circ$ (Figure \ref{fig:th87bp5m3p5:cf:100:10keV}):
a plateau region (now with a definite secondary peak)
is present for an intermediate range of injection energy, and
a power law decrease at higher energies. The slope is independent
of injection energy.
However, for $E_i=5$~keV and 10~keV there is a single
power law decrease, over 5 orders of magnitude,
from the injection energy up to about 100~keV. There is some
indication that the fall off is more rapid above 100~keV.

\begin{figure}
\epsscale{.80}
\plotone{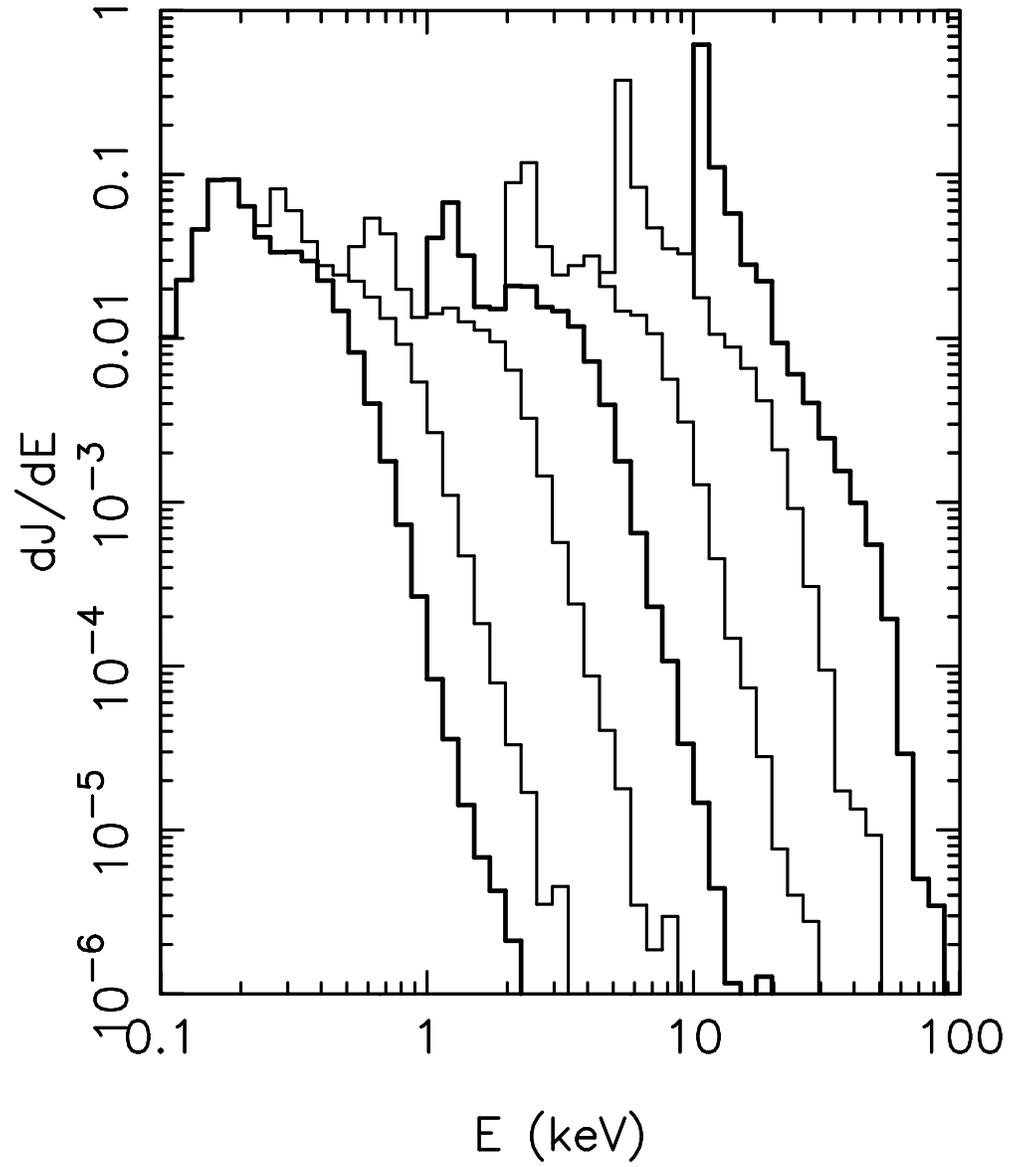}
\caption{As for Figure \ref{fig:th87bp5m3p5:cf:100:10keV}
for a shock similar to Case A, but with $\theta_{Bn}=85^\circ$
\label{fig:th85bp5m3p5:cf:100:10keV}}
\end{figure}

\begin{figure}
\epsscale{.80}
\plotone{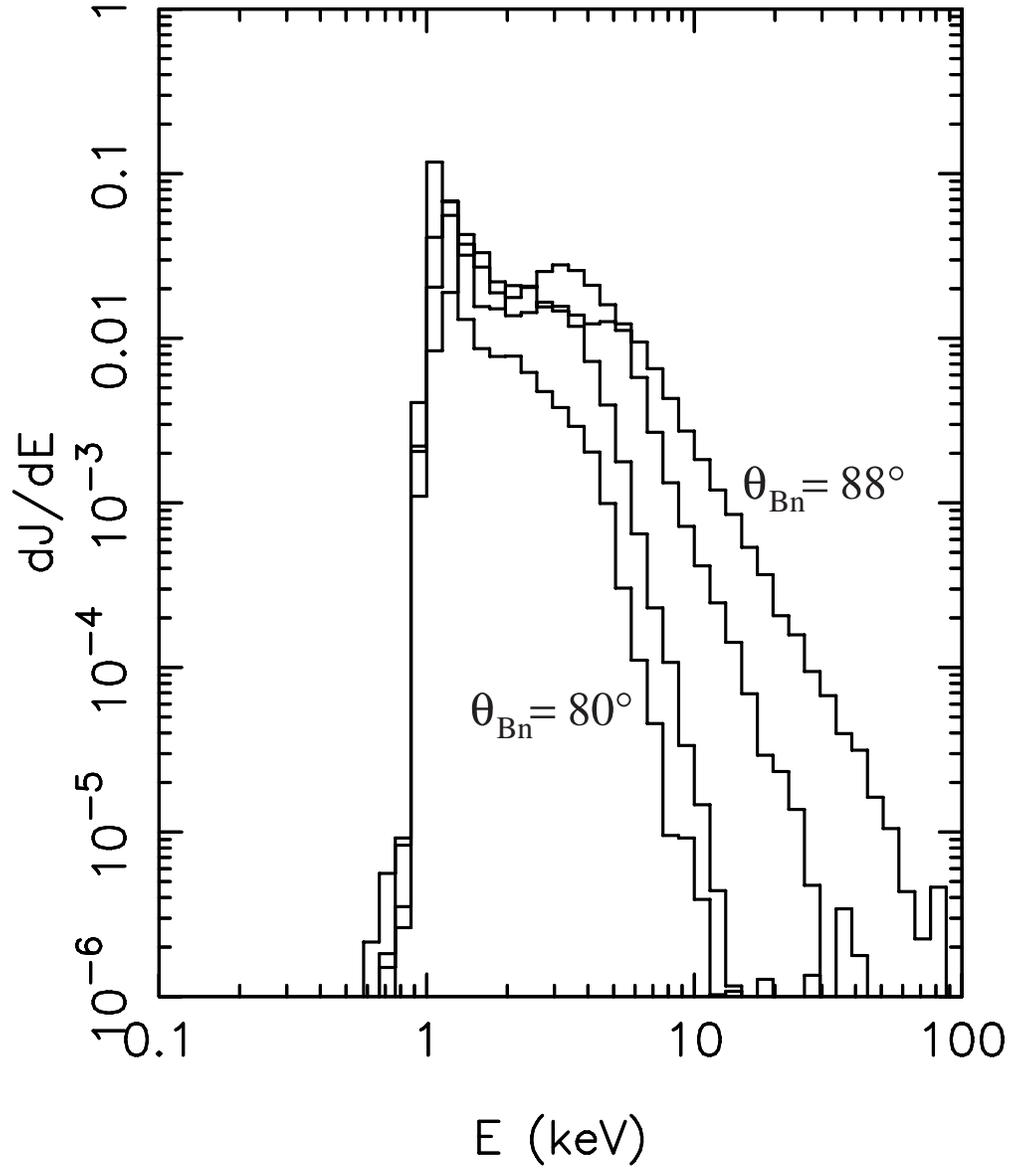}
\caption{Energy spectra for a fixed injection energy of 1~keV for
a set of shocks similar to Case A, but with $\theta_{Bn}=80^\circ$, $85^\circ$,
$87^\circ$, and $88^\circ$.
\label{fig:1keV:cf:th80:th88}}
\end{figure}

Figure
\ref{fig:th85bp5m3p5:cf:100:10keV} shows the spectra
for $E_i$ from 100~eV to 10~keV for a similar simulation with 
a lower value of $\theta_{Bn}=85^\circ$. Once again the spectra 
show broadly the same features as for $\theta_{Bn}=87^\circ$,
although the maximum energies achieved are lower. The
effects of $\theta_{Bn}$ on the fall off slope and flux levels
are illustrated in Figure \ref{fig:1keV:cf:th80:th88}, which
shows the energy spectra for $E_i=1$~keV for $\theta_{Bn}=
80^\circ$, 85$^\circ$, 87$^\circ$, and 88$^\circ$.
The effect of increasing $\theta_{Bn}$ is to increase the
maximum energy at a given flux level, to emphasize the
appearance of the plateau/secondary peak region of the
spectrum, and to decrease the slope of power law decrease.
The power slope varies from -3.1 at $\theta_{Bn}=88^\circ$
to -7.0 at $\theta_{Bn}=80^\circ$.
Another key feature is that the flux levels in the
intermediate portion of the spectrum are all approximately
equal in the range $\theta_{Bn}=85^\circ - 88^\circ$
(and presumably above).
Even at $\theta_{Bn}=80^\circ$, the flux levels are only
down by a factor $0.1 - 0.5$. This indicates that the
acceleration mechanism in the intermediate range operates
over a reasonably wide range of $\theta_{Bn}$, at least
when compared to adiabatic reflection theory, which is
often characterized as most effective above $\theta_{Bn}=88^\circ$.

\begin{figure}
\epsscale{.80}
\plotone{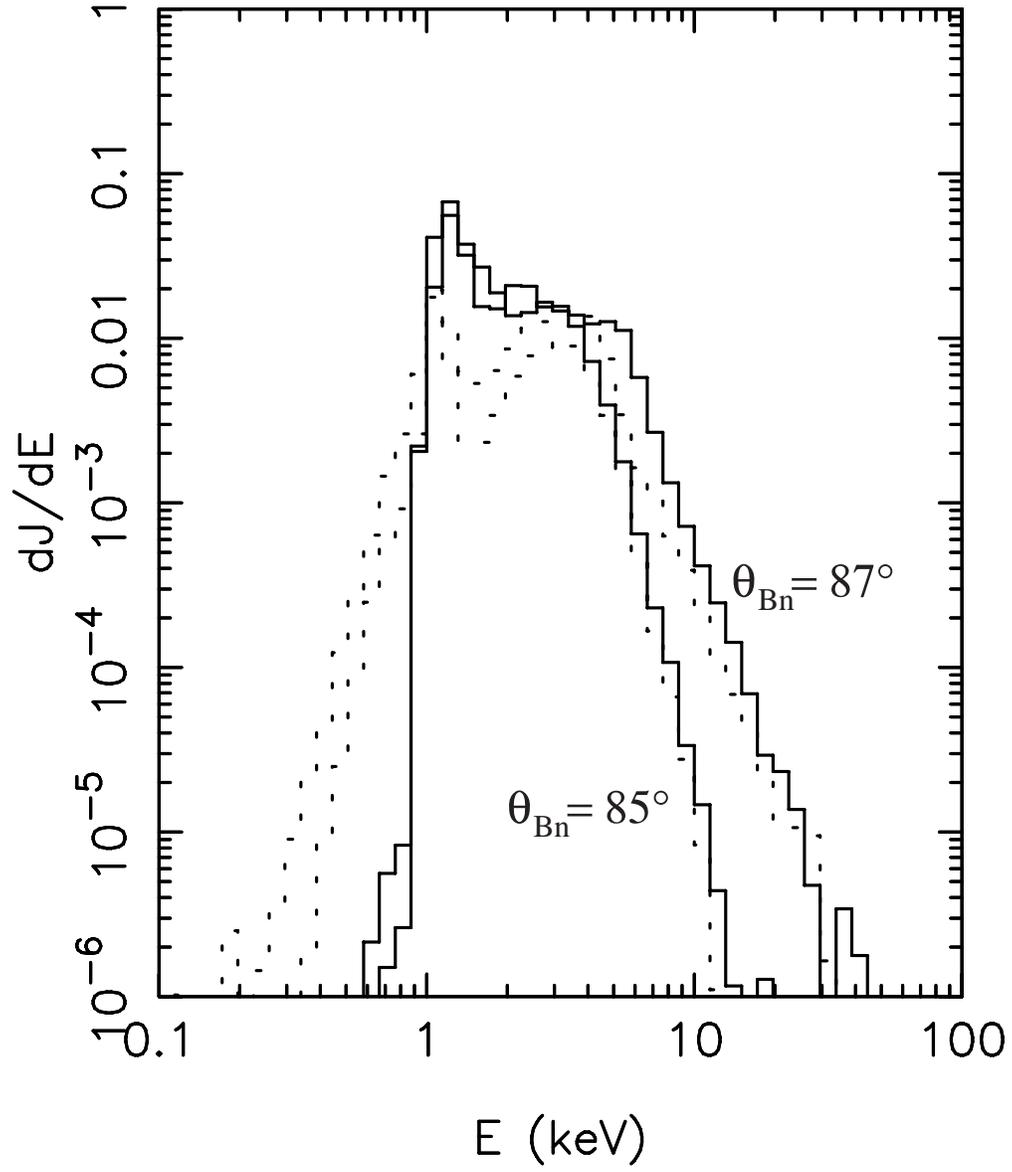}
\caption{Upstream (solid) and downstream (dotted) energy spectra
for fixed injection energy of 1~keV for two shocks: Case A
with $\theta_{Bn}=87^\circ$, and a similar shock with $\theta_{Bn}=85^\circ$.
\label{fig:1keV:updown:th85:th88}}
\end{figure}

\begin{figure}
\epsscale{.80}
\plotone{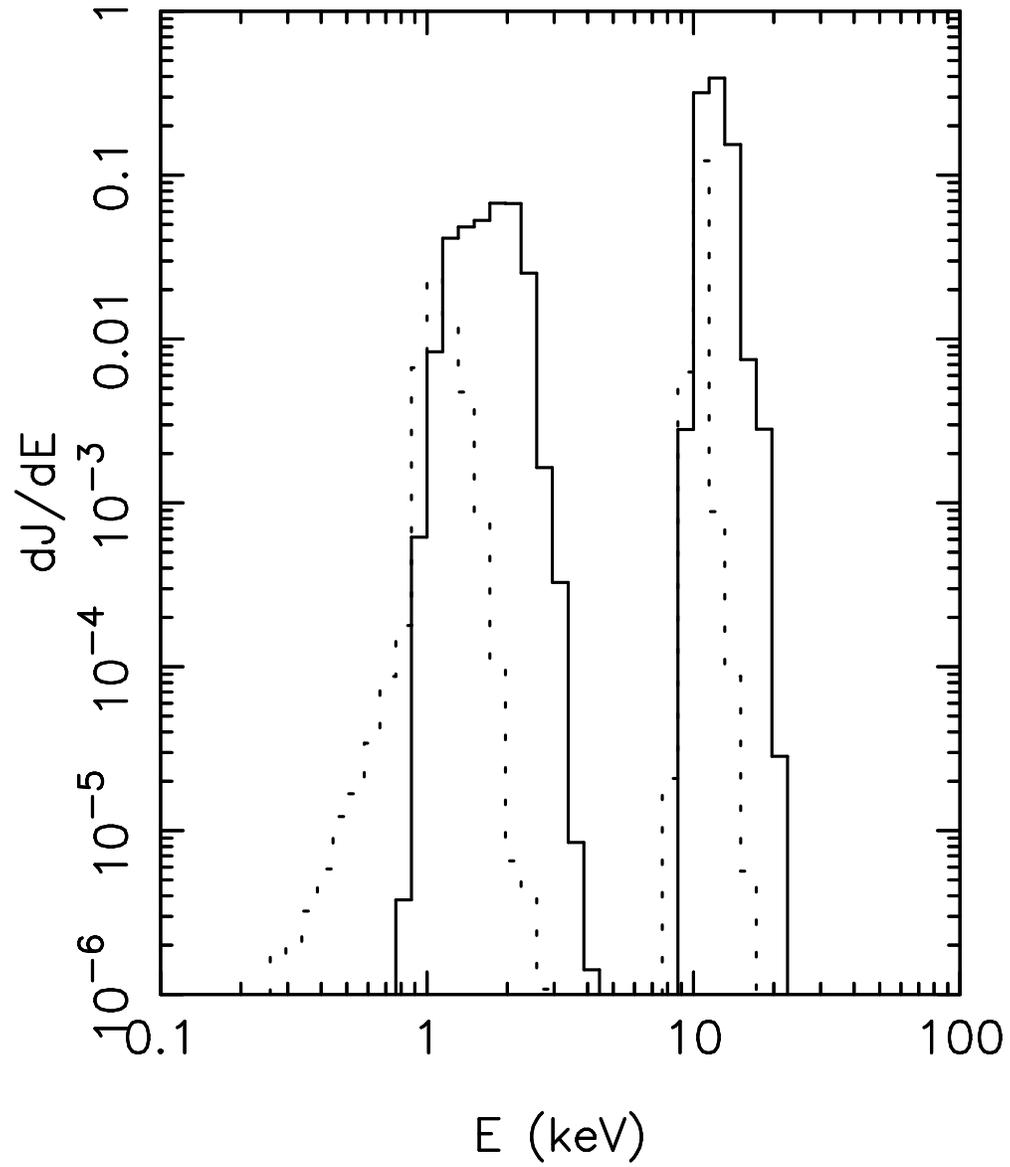}
\caption{Upstream (solid) and downstream (dotted) energy spectra
for Case B (without shock surface structuring) for injection
energies of 1~keV  and 10~keV.
\label{fig:1keV:updown:th87oop}}
\end{figure}

Finally, we have so far concentrated on the results for
electrons collected upstream, since the reflected population
is the most studied case at the Earth's bow shock in
terms of observations and implications for waves
in the foreshock. Figure \ref{fig:1keV:updown:th85:th88} compares
the upstream and downstream spectra for an injection
energy of 1~keV for $\theta_{Bn}=88^\circ$ and
$\theta_{Bn}=85^\circ$. Especially striking is the fact
that the upstream and downstream spectra for
the region of power law fall off, beyond the
the plateau region, are almost exactly the same in
terms of level and slope. This is a very general
property for all the cases we have studied where the
upstream magnetic field lies in the simulation plane,
i.e., where the shock has dynamic rippled structuring.
The greatest difference between the upstream and
downstream spectra is in the range $1-3$~keV; there
are greater differences as $\theta_{Bn}$ decreases.
For example, at $\theta_{Bn}=85^\circ$, there is
no plateau region in the downstream spectrum, whereas
the upstream spectrum shows a plateau region at
a high flux level. The similarity of the upstream and
downstream spectra can be shown to be directly linked
to the presence of dynamic shock structure. Figure
\ref{fig:1keV:updown:th87oop} shows the comparison
between the upstream and downstream spectra
(for $E_i=1$ and $10$~keV) for
a shock simulation where the structuring along
the shock surface is suppressed (Case B). There are
major differences between the level and form of the
upstream and downstream spectra. The downstream
spectra shows hardly any energization, with
the upstream flux at a very much higher level at all
energies above the injection energy. As discussed above,
the upstream spectra themselves have a steep fall off
with energy.

\section{Summary} \label{sect:summary}

The work presented here extends earlier work studying fast Fermi
shock acceleration of electrons using self-consistently simulated
shock fields. A two dimensional hybrid shock simulation has been used,
which shows dynamic rippled structure across the shock surface
and along the magnetic field direction. This structuring produces
a radically different behavior compared with a one dimensional
shock. Our results indicate that suprathermal electrons can be scattered
within the shock transition even without strong electron scale fluctuations.
For a shock with $M_A=5$ and
using test particle electrons with injection energies
in the range $100 - 10$~keV, we find energy spectra of differential flux to
have a ``plateau'' region (either flat or shallow sloped) in an
intermediate energy range up to $2 - 5$ times the injection energy, with
a larger range at lower energies. Above the plateau region of the
spectrum, there is an inverse power law form, the slope of
which increases as $\theta_{Bn}$ decreases. Above a certain
energy the downstream and upstream spectra are effectively
the same. The accelerated fluxes are present over an extended
range of $\theta_{Bn}$, not just immediately close to $90^\circ$,
as found for adiabatic reflection theory. All these principle
results are in general agreement with observations. As such, these 
results have implications for models of electron acceleration
at planetary bow shocks, interplanetary shocks (and similar
other astrophysical shocks) and also for models of nonthermal
radio-emission which depend on accelerated electrons. More detailed
comparisons between the simulations and
observations will use an initial electron
distribution, rather than the mono-energetic release used here.

The similarity of the downstream and upstream spectra above
some energy indicates that the electron motion within
the shock transition is strongly affected by a scattering
process, which is not surprising given the level of fluctuation
at the shock surface. At lower energies, however, there is
a clear signature of the adiabatic reflection process. This 
suggests that both processes play a role: an initial energization
is produced by magnetic mirroring, but scattering on magnetic
fluctuations keeps particles within the shock transition, and
allows stochastic acceleration. Particles with suitable energy
and pitch angle at the edges of the shock transition region
can then escape either upstream or downstream. A scenario with
strong scattering in the shock transition will produce
two effects: an increase in accelerated electron flux at
the shock itself, i.e., a ``shock spike event'' as observed
at the Earth's bow shock \citep{gosling++1989} and
interplanetary shocks \citep{tsurutani+lin1985}; and near
isotropic
downstream distributions with the upstream ``reflected''
population emerging from the downstream distribution
in the shock ramp, as observed at the bow shock \citep{gosling++1989}.

The use of a hybrid simulation to provide the electric and magnetic
fields means that electron scale (in the simulation frame)
fluctuations are not included. Consequently, our results may not
be appropriate for shocks which have strong electron heating and
therefore strong electron scale turbulence. This is not generally
the case for planetary and interplanetary shocks. The validity of
neglecting such electron scale fluctuations can only be validated
by (at least two dimensional) full particle simulations with careful
use of appropriate simulation parameters (see discussion in Section
\ref{sect:introduction}). However, even if, for some given shock,
there is a non-negligible effect from electron scale turbulence, then
the process described in this paper will still operate, provided
that there is dynamic rippled structuring of the shock front. In this
case, the relative importance of the two processes will have to be
evaluated.

Our key results, such as inverse power law energy spectra and extended
range of effective $\theta_{Bn}$,
are due to the presence of shock surface fluctuations, which
have, in turn, been shown to be present only if the shock Mach number
is high enough. This provides a key observational test, since
interplanetary shocks cover a wide range of Mach number. Also, 
Type II solar radio bursts show considerable variability, and this
may be due to the associated interplanetary shock changing Mach number
as it propagates through an inhomogeneous solar wind
\citep[c.f.,][]{knock++2003}. Our results show that a low Mach number
shock should exhibit the features of scatter free adiabatic
reflection theory: upstream maximum energization 
close to $\theta_{Bn}=90^\circ$ with a limited
energy range, and anisotropic downstream distributions
with little energization. However, this picture maybe
modified if the shock is propagating through turbulence. It is
possible that
fluctuations of the appropriate scale lengths are amplified
at the shock, thereby producing scattering in the
same way as the surface fluctuations at the high Mach number shocks
simulated here.

We have presented results for mono-energetic injection,
thereby demonstrating  the overall
operation and behavior of the process and the range of parameters
for which it is effective. Future work will present the evolution
of the energetic electron distribution function across the shock,
and the spectra and anisotropies for model input distribution functions,
thereby facilitating comparisons with observations



\acknowledgments

We are grateful to R.~E.~Lowe who contributed to software
development.






\begin{thebibliography}{30}
\expandafter\ifx\csname natexlab\endcsname\relax\def\natexlab#1{#1}\fi

\bibitem[{Anderson {et~al.}(1979)Anderson, Lin, Martel, Lin, Parks, \&
  R{\`e}me}]{anderson++1979}
Anderson, K.~A., Lin, R.~P., Martel, F., Lin, C.~S., Parks, G.~K., \& R{\`e}me,
  H. 1979, \grl, 6, 401

\bibitem[{{Bessho} \& {Ohsawa}(2002)}]{bessho+ohsawa2002}
{Bessho}, N., \& {Ohsawa}, Y. 2002, Physics of Plasmas, 9, 979

\bibitem[{Burgess(1995)}]{burgess1995}
Burgess, D. 1995, in Introduction to Space Physics, M. G. Kivelson and C. T.
  Russell (eds) (Cambridge: Cambridge University Press), 129--163

\bibitem[{{Cairns}(1987)}]{cairns1987}
{Cairns}, I.~H. 1987, \jgr, 92, 2315

\bibitem[{{Cargill} \& {Papadopoulos}(1988)}]{cargill+papadopoulos1988}
{Cargill}, P.~J., \& {Papadopoulos}, K. 1988, \apjl, 329, L29

\bibitem[{Fitzenreiter {et~al.}(1990)Fitzenreiter, Scudder, \&
  Klimas}]{fitzenreiter++1990}
Fitzenreiter, R.~J., Scudder, J.~D., \& Klimas, A.~J. 1990, \jgr, 95, 4155

\bibitem[{Gosling {et~al.}(1989)Gosling, Thomsen, Bame, \&
  Russell}]{gosling++1989}
Gosling, J.~T., Thomsen, M.~F., Bame, S.~J., \& Russell, C.~T. 1989, \jgr, 94,
  10011

\bibitem[{{Holman} \& {Pesses}(1983)}]{holman+pesses1983}
{Holman}, G.~D., \& {Pesses}, M.~E. 1983, \apj, 267, 837

\bibitem[{{Knock} {et~al.}(2003){Knock}, {Cairns}, {Robinson}, \&
  {Kuncic}}]{knock++2003}
{Knock}, S.~A., {Cairns}, I.~H., {Robinson}, P.~A., \& {Kuncic}, Z. 2003, \jgr,
  108, 6

\bibitem[{{Krauss-Varban}(1994)}]{kraussvarban1994}
{Krauss-Varban}, D. 1994, J. Geophys. Res., 99, 2537

\bibitem[{{Krauss-Varban} \& Burgess(1991)}]{kraussvarban+burgess1991}
{Krauss-Varban}, D., \& Burgess, D. 1991, J. Geophys. Res., 96, 143

\bibitem[{{Krauss-Varban} {et~al.}(1989){Krauss-Varban}, Burgess, \&
  Wu}]{kraussvarban++1989}
{Krauss-Varban}, D., Burgess, D., \& Wu, C.~S. 1989, J. Geophys. Res., 94,
  15089

\bibitem[{{Krauss-Varban} \& Wu(1989)}]{kraussvarban+wu1989}
{Krauss-Varban}, D., \& Wu, C.~S. 1989, J. Geophys. Res., 94, 15367

\bibitem[{{Kuncic} {et~al.}(2004){Kuncic}, {Cairns}, \& {Knock}}]{kuncic++2004}
{Kuncic}, Z., {Cairns}, I.~H., \& {Knock}, S.~A. 2004, \jgr, 109, A02108

\bibitem[{{Lemb{\`e}ge} \& {Savoini}(2002)}]{lembege+savoini2002}
{Lemb{\`e}ge}, B., \& {Savoini}, P. 2002, Journal of Geophysical Research
  (Space Physics), 107

\bibitem[{Leroy \& Mangeney(1984)}]{leroy+mangeney1984}
Leroy, M.~M., \& Mangeney, A. 1984, Annales Geophys., 2, 449

\bibitem[{Lowe \& Burgess(2003)}]{lowe+burgess2003}
Lowe, R.~E., \& Burgess, D. 2003, Ann.\ Geophys., 21, 671

\bibitem[{Matthews(1994)}]{matthews1994}
Matthews, A.~P. 1994, J.\ Comp.\ Phys., 112, 102

\bibitem[{{McKean} {et~al.}(1995){McKean}, {Omidi}, \&
  {Krauss-Varban}}]{mckean++1995}
{McKean}, M.~E., {Omidi}, N., \& {Krauss-Varban}, D. 1995, J. Geophys. Res.,
  100, 3427

\bibitem[{{Savoini} \& {Lemb{\` e}ge}(2001)}]{savoini+lembege2001}
{Savoini}, P., \& {Lemb{\` e}ge}, B. 2001, \jgr, 106, 12975

\bibitem[{{Schmitz} {et~al.}(2002){Schmitz}, {Chapman}, \&
  {Dendy}}]{schmitz++2002}
{Schmitz}, H., {Chapman}, S.~C., \& {Dendy}, R.~O. 2002, \apj, 579, 327

\bibitem[{{Scholer} {et~al.}(2003){Scholer}, {Shinohara}, \&
  {Matsukiyo}}]{scholer++2003}
{Scholer}, M., {Shinohara}, I., \& {Matsukiyo}, S. 2003, \jgr, 108, 4

\bibitem[{{Simnett} {et~al.}(2005){Simnett}, {Sakai}, \&
  {Forsyth}}]{simnett++2005}
{Simnett}, G.~M., {Sakai}, J.-I., \& {Forsyth}, R.~J. 2005, \aap, 440, 759

\bibitem[{Thomsen(1968)}]{thomson1968}
Thomsen, W.~E. 1968, Comput. J., 10, 417

\bibitem[{{Tsurutani} \& {Lin}(1985)}]{tsurutani+lin1985}
{Tsurutani}, B.~T., \& {Lin}, R.~P. 1985, \jgr, 90, 1

\bibitem[{{Vandas}(1995)}]{vandas1995b}
{Vandas}, M. 1995, \jgr, 100, 23499

\bibitem[{{Vandas}(2001)}]{vandas2001}
---. 2001, \jgr, 106, 1859

\bibitem[{{Vandas} \& {Karlick{\'y}}(2000)}]{vandas+karlicky2000}
{Vandas}, M., \& {Karlick{\'y}}, M. 2000, \solphys, 197, 85

\bibitem[{Winske \& Quest(1988)}]{winske+quest1988}
Winske, D., \& Quest, K.~B. 1988, J. Geophys. Res., 93, 9681

\bibitem[{Wu(1984)}]{wu1984}
Wu, C.~S. 1984, \jgr, 89, 8857

\end{thebibliography}

%

\clearpage



\end{document}